\documentclass[12pt]{article}
\begin{document}
\title{RAMIFICATIONS OF NON COMMUTATIVE SPACETIME}
\author{B.G. Sidharth\\
Centre for Applicable Mathematics \& Computer Sciences\\
B.M. Birla Science Centre, Adarsh Nagar, Hyderabad - 500 063 (India)}
\date{}
\maketitle
\begin{abstract}
We review Cantorian and Non Commutative Spacetime, work which has occupied El Naschie in the past several years. These concepts are now the subject of intense research, thanks to Quantum Gravity, Quantum Super String Theory and a few other approaches. It now appears that we are on the verge of a breakthrough in finding solutions to longstanding problems, like the unification of gravitation with other fundamental interactions and the question of the mass spectrum, another recent area of El Naschie's work.
\end{abstract}
\section{Introduction}
It is now a cliche that the two great intellectual pillars of the twentieth century, viz., General Relativity or, more generally, Gravitation and Quantum Mechanics have stood apart, stubbornly defying attempts at unifying them. As Wheeler \cite{MWT} noted, the problem finally boils down to the introduction of the Quantum Mechanical concept of spin half into Classical Theory and the classical concept of curvature into Quantum Theory. Curiously enough the two pillars of General Relativity and Quantum Theory stand on a common ground: Together they use the concept of a differentiable space time manifold, be it the Reimannian space time of General Relativity or the Minkowski space time of Relativistic Quantum Theory or Quantum Field Theory. However more recent work be it in Quantum Gravity or in Quantum Super String Theory, has hinted at a minimum space time cut off \cite{Amati,Garay,MUP}. This alters the age cold concept of a smooth space time. In recent years there has been quite some work by different authors such as Ord, Nottale, El Naschie, the author and others in this relatively new field of non differentiable space time \cite{Ord,Ord2,Nottale,El1,El2,El3,CU} and several references therein. One of the very fruitful concepts put forward in these pathbreaking efforts has been El Naschie's concept of the Cantorian space time. Atlast some schoalrs  are beginning to realize the fractal nature of space time. Once we break out of the smooth spacetime mindset, many exciting possibilities open up, including a unification of gravitation with other fundamental interactions, and also the possibility of solving the elusive problem of the mass spectrum, as  has been deduced by El Naschie for example \cite{El4}.\\
We will briefly survey some of these efforts and indicate how the solution of longstanding problems are now within sight.
\section{Non Commutative Spacetime}
We start with a Quantum Mechanical description. In Quantum Theory, it is well known that spacetime points pose a major difficulty. By the Uncertainity principle, if we go down to such points, we will end up with infinite momenta and energies. This situation comes up clearly in the case of the Quantum Mechanical treatment of the electron \cite{Dirac}. We encounter electrons with the velocity of light and rapid oscillations termed zitterbewegung. At that time Dirac realised this problem and correctly explained the anamoly by noting that only observations averaged over small intervals at the Compton scale are physically meaningful. Within the Compton scale there are well known non local effects. This is a major departure from classical considerations where we use spacetime points.\\
Nevertheless Quantum Theory has continued using classical space time as a background. This has lead, for example to the problem of infinities and resonances found by Poincare much earlier in classical theory \cite{Rohrlick}. For example in Classical Electrodynamics, the relativistic generalisation of the well known Lorentz equation is the Lorentz-Dirac equation. This equation shows unphysical characteristics, such as third derivatives with respect to time, run away solutions with infinite energy and so on \cite{Rohrlick,IAAD}. The infinite energy in this case is attributed to the infinite energy which an electron, treated as a spherical shell, acquires when the radius of the shell tends to zero. But it is remarkable that even in Classical Electrodynamics the infinities and non local effects are confined to the same Quantum Mechanical Compton scale \cite{IAAD}.\\
One way in which Quantum Field Theory has dodged these unpleasant features is by using the renormalization technique. To put it briefly, the physical quantity we observe, for example the mass or charge consist of two components: One is the ``actual'' or bare value and the other is a value due to the above effects. What is important ultimately is the physical quantity we observe. So if we can find a model in which the bare value and the infinite value somehow add up to give the correct physical value in the limit, then the theory is renormalizable, because in any case what happens within the small intervals is ultimately of no consequence \cite{Hooft,Klauder}. Dirac himself was critical of this approach and predicted that one day it would be proved wrong \cite{Sachs}. He observed, ``I am inclined to suspect that the renormalization theory is something that will not survive in the future, and that the remarkable agreement between its results and experiment should be looked on as a fluke.''\\
In parallel a number of scholars such as Snyder had tried to work out theories which recognized minimum intervals as a way to eliminate the infinities \cite{Snyder1,Snyder2,CU}.This work did not find favour for many years.\\
However relatively recent developments from two separate directions, that of Quantum Gravity and Quantum Super String theory have lead back to the notion of a minimum space time scale \cite{Garay,Amati,MUP}: The Planck Scale ($10^{-33}cms$ and $10^{-43}secs$) which is the Compton scale for a particle with a Planck mass, $10^{-5}gms$. Interestingly Max Planck himself had noted that the Planck scale is made up of a combination of the fundamental constants and therefore must be fundamental. In recent years there has been a growing body of literature which argues that the Planck scale is indeed such a fundamental minimum scale.\\
The introduction of such a minimum scale, instead of space time points leads to consequences far beyond the classical description. Space time now becomes non commutative \cite{AFDB,Rev}, the Uncertainity principle takes on an extra term, sometimes called the duality term \cite{Rev,BGSNC2002} and so on. Specifically, we have, 
$$[x,y] = (\imath a^2/\hbar) L_z, [t,x] = (\imath a^2/\hbar c)M_x,$$
\begin{equation}
[y,z] = (\imath z^2/\hbar)L_x, [t,y] = (\imath a^2/\hbar c) M_y,\label{ea}
\end{equation}
$$[x,x] = (\imath a^2\hbar)L_y, [t,z] = (\imath a^2/\hbar c)M_z,$$
$$[x,p_x] = \imath \hbar [1 + (a/\hbar)^2 p^2_x];$$
$$[t,p_t] = \imath \hbar [1 - (a/\hbar c)^2 p^2_t];$$
\begin{equation}
[x, p_y] = [y, p_x] = \imath \hbar (a/\hbar)^2 p_xp_y;\label{eb}
\end{equation}
$$[x,p_t] = c^2 [p_x, t] = \imath \hbar (a/\hbar)^2 p_xp_t;etc.$$
Non relativistic Quantum Theory follows from (\ref{ea}) if $a^2$ is neglected, though spacetime is now commutative as in the usual theory.
The modification to the Uncertainity Principle is seen in equation (\ref{eb}) when terms $\sim 0 (a^2)$ are retained.
All this also averts, what Wheeler had called the greatest crisis of Physics, namely the space time singularity of the point spacetime theory, indeed as the point no longer exists.\\
It can be argued that the space time generally used in Classical theory and Quantum Theory is an approximation, which smoothens out an underpinning chaotic behavior at the Planck scale \cite{BGSPSP,SUSY}. Indeed as Wheeler himself observed, ``No prediction of space time, therefore no meaning for space time is the verdict.... That object which is central to all of classical general relativity, the four dimensional space time geometry, simply does not exist, except in a classical approximation.... One has to forego that view of nature in which every event, past, present, or future, occupies its preordained position in a grand catalog called ``space time''....''\cite{MWT}. This approximate classical space time is quasi- changeless or stationary and time is reversible, as indeed is evident from the equations of motion both in Classical Physics and Quantum Physics. But when we go beyond this approximation, to the stochastic description at the Planck scale \cite{BGSPSP} time is no longer reversible. We make the leap from the age old concept of ``being'' to the concept of ``becoming'' at least at this scale.\\
One of the criticisms put forward against the Planck Scale Phenomena is that these effects are beyond experimental verification for a long time to come. But let us analyse this further. The Compton scale which we encounter in the physical world has an underpinning of some $n = 10^{40}$ transient Planck particles. However Planck scale phenomena are moderated, and we have, as in a diffusion process,
\begin{equation}
l = \sqrt{n} l_P\label{e1}
\end{equation}
\begin{equation}
m = m_P / \sqrt{n}\label{e2}
\end{equation}
where $l$ and $m$ are the Compton wavelength and mass of a typical elementary particle and $l_P$ and $m_P$ are the Planck length and the Planck mass. An equation identical to (\ref{e1}) holds for the Compton time also.\\
(\ref{e1}) and (\ref{e2}) are not mere numerical accidents - they can be deduced as mentioned from a diffusion process \cite{BGSPSP}. The $\sqrt{n}$ in the equations is indicative of a Brownian process. For example, in a random walk of $n$ steps, each of length $l$, the total distance covered would be of the order of $\sqrt{n}l$.\\
Infact we can go one step further. Remembering that there are a total of  $N = 10^{80}$ elementary particles, the entire universe shows up as $n \times N = 10^{120}$ Planck scale oscillators. Using the fact that the $r$th energy led for the Harmonic oscillators is given by $\sqrt{r} \hbar \omega$ for large $r$, \cite{BGSFFP5} it follows that the total energy of these Planck scale oscillators would be $\sqrt{nN} m_Pc^2$, which correctly gives the mass of the universe itself. That is, the universe is a normal mode of these Planck scale oscillators.\\
We have already remarked that the Quantum commutators are present in (\ref{ea}), if we neglect terms of the order of $a^2$. In particular, taking $a$ to be the Compton scale, it has been shown that we can recover from (\ref{ea}) and (\ref{eb}) the Dirac equation itself\cite{CU}. This again is not surprising because the non commutativity in (\ref{ea}) and (\ref{eb}) can be shown to represent spin even from the classical viewpoint \cite{Zak}. Interestingly, at the Compton scale it has been shown that the Quantum coordinates coincide with the complex coordinates of a classical Kerr-Newman Black Hole with radius of the order of the Compton wavelength. Indeed it has been known for a long time that the classical Kerr-Newman metric reproduces the field of the electron including the purely Quantum Mechanical gyro magnetic ratio $g = 2$. What has been inexplicable is the fact that there is a naked singularity or equivalently complex coordinates. This infact is a direct consequence of the non commutativity \cite{CU}. In other words once the non commutative or fuzzy nature of spacetime is taken into consideration, corresponding to averages over zitterbewegung in Realtivistic Quantum Theory, the naked singularity disappears and the electron can be represented by the Kerr-Newman metric. These conclusions have since been confirmed by Nottale \cite{Nottale1}. Already this Kerr-Newman characterization of the Quantum Mechanical electron points to the long sought after linkage between gravitation and electromagnetism \cite{BGSGrav}. It can be shown formally that this is so \cite{NCB116B,AFDB,BGSNC2002}. Infact arising from (\ref{ea}) there is the covariant derivative
\begin{equation}
\partial_\mu \to \partial_\mu - \Gamma^\sigma_{\mu^\sigma}\label{eX}
\end{equation}
The second term on the right of (\ref{eX}) represents the electromagnetic potential, and surprisingly coincides with the original Weyl formulation of electromagnetism.\\
Surprising as it may seem there is a cosmological scheme which follows from these considerations. It can be understood on the basis of the fact that in the minimum time interval $\tau$ at the Compton scale ,$\sqrt{N}$ particles are created fluctuationally from the Quantum vacuum, where $N$ is the number of particles present at that epoch \cite{IJMPA,IJTP,BGSCosfluc}. This cosmology successfully predicted dark energy and an accelerating ever expanding universe since confirmed by observation \cite{Perl,Kirsh}.\\
All this can also be shown to explain some hitherto inexplicable coincidences noted by Weyl, Eddington, Dirac and Weinberg. These are relations like
\begin{equation}
R = \sqrt{N} l\label{e3}
\end{equation}
\begin{equation}
\frac{e^2}{Gm^2} = \sqrt{N}\label{e4}
\end{equation}
\begin{equation}
m = \left(\frac{\hbar^2 H}{Gc}\right)^{1/3}\label{e5}
\end{equation}
and others. $R$ is the radius of the universe, $e$ is the electron charge, $G$ the universal constant of gravitation, $c$ the velocity of light, $\hbar$ the reduced Planck constant and $H$ the Hubble constant. It is easy and even unscientific to dismiss such equations as accidents. Dirac himself realised that equations (\ref{e3}) and (\ref{e4}) for example could have a cosmological significance \cite{Narlikar}. There was an inconsistency in his otherwise beautiful cosmology. More recently it has been shown by the author that these coincidental equations (\ref{e5})  can be deduced, on the basis of the cosmology mentioned above. It may be mentioned that Weinberg had termed the equation (\ref{e5}) as being mysterious because it relates a large scale parameter like $H$, the Hubble constant to microphysical constants \cite{Weinberg}. This points to a Machian or ``co-related'' universe which is not surprising because the universe as we have just seen, is a normal mode of Planck oscillators.\\
In this consistent  scheme, the universe is created out of a sub stratum Quantum vaccuum or dark energy, in a phase transition at the Planck scale \cite{Cosmos}. Herein are the very first seeds for all the complex structures of the universe.
\section{A Mass Spectrum}   
We can extend these considerations to generate a mass spectrum, a problem that has fascinated El Naschie and for which his model gives a solution \cite{El4}. We now use the model of three oscillators (typically for the three quarks), discussed in detail in references \cite{MPLA1,MPLA2,CU,FFPLobanov}. We  use the fact that for such an oscillator (resembling a triatomic molecule) \cite{Goodstein} the frequencies are given by
\begin{equation}
\omega = \sqrt{k/m}, 2  \sqrt{k/m}\label{ex}
\end{equation}
for one and two such oscillators where $\hbar \omega = mc^2, m$ being the mass.\\
For reasons discussed in detail  in the references, we take the higher frequency or mass to represent the pion $m_\pi$. In this connection, it may be mentioned that the $\pi^0$ meson has been shown to be a bound state of an electron and positron in the discrete spacetime theory \cite{CU}, borne out by its decay mode. This was a starting point for El Naschie in his mass spectrum  model. Then from (\ref{ex}) we derive the mass spectrum from the oscillator energy levels viz.,
\begin{equation}
m \approx (n + \frac{1}{2}) m_\pi , m \approx (2n+1) m_\pi, n = 0,1,2,\cdots\label{ey}
\end{equation}
It is remarkable that the formula (\ref{ey}) generates a whole range of some $50$ mesons and baryons including the well known particles from the particle data group tables (Cf.refs.\cite{Hagiwar}), starting from the $K$meson ($480 meV$ approximately) through several other mesons including the $2 \chi, D, f, \rho, \Theta$ and so on as also the very massive $\gamma (1s), \chi 1p, \gamma 2s, \gamma 3s, \gamma 4s, \gamma 10860, B, J/ (\psi 1s)$ and so on, going right up to values of $n$ near $80$ giving the heaviest of the particles and clusters of masses nearby. Even in the approximation (\ref{ey}) where several degrees of freedom and other details have been excluded, the agreement is within a few percent of the actual values. Interestingly the latest particle $D_5(2317)$ \cite{Nature} also follows from (\ref{ey}), being apprxomately $17m_\pi$. Incidentally  El Naschie's mass spectrum can also start with the model of spring connected oscillators \cite{El4}.

\end{document}